\documentclass[preprint,showpacs,showkeys,pre]{revtex4}
\usepackage{amssymb}
\usepackage{amsmath}
\usepackage{epsfig}
\usepackage{psfrag}
\usepackage{subfigure}

\newcommand{\etal}{\textit{et al.\ }}
\newcommand{\ie}{i.e.\ }
\newcommand{\eg}{e.g.\ }
\newcommand{\etc}{etc.\ }
\newcommand{\R}{\mathbb{R}}
\newcommand{\C}{\mathbb{C}}
\newcommand{\e}{\mathrm{e}}

\begin{document}

\title{Stabilizing unstable periodic orbits in the Lorenz equations
  using time-delayed feedback control}
\author{Claire M. Postlethwaite}
\email{c-postlethwaite@northwestern.edu}
\author{Mary Silber}
\affiliation{Engineering Sciences and Applied Mathematics,
  Northwestern University, Evanston, IL, 60208, USA}
\date{\today}

\begin{abstract}
For many years it was believed that an unstable periodic orbit with an
odd number of real Floquet multipliers greater than unity cannot be
stabilized by the time-delayed feedback control mechanism of Pyragus.
A recent paper by Fiedler~\etal\cite{Fie06} uses the normal form of a
subcritical Hopf bifurcation to give a counterexample to this theorem.
Using the Lorenz equations as an example, we demonstrate that the
stabilization mechanism identified by Fiedler~\etal for the Hopf
normal form can also apply to unstable periodic orbits created by
subcritical Hopf bifurcations in higher-dimensional dynamical
systems. Our analysis focuses on a particular codimension-two
bifurcation that captures the stabilization mechanism in the Hopf
normal form example, and  we show that the same codimension-two
bifurcation is present in the Lorenz equations with appropriately
chosen Pyragus-type time-delayed feedback. This example suggests a
possible strategy for choosing the feedback gain matrix in Pyragus
control of unstable periodic orbits that arise from a subcritical Hopf
bifurcation of a stable equilibrium. In particular, our choice of
feedback gain matrix is informed by the Fiedler~\etal example, and
it works over a broad range of parameters, despite the fact that a
center-manifold reduction of the higher-dimensional problem does not
lead to their model problem.
\end{abstract}


\maketitle

\section{Introduction}

Time-delayed feedback control has been used as a method of stabilizing
unstable periodic orbits (UPOs) or spatially extended patterns by a
number of authors. The method of Pyragus~\cite{Pyr92}, sometimes	
called `time-delayed autosynchronization' (TDAS), has attracted much
attention. Here, the feedback $F$ is proportional to the difference
between the current and a past state of the system. That is,
$F=K(x(t-\tau)-x(t))$ where $x(t)$ is some state vector, $\tau$ is the
period of the targeted UPO and $K$ is a feedback gain
matrix. Advantages of this method include the following. First, since
the feedback vanishes on any orbit with period $\tau$, the targeted
UPO is still a solution of the system with feedback. Control is
therefore achieved in a non-invasive manner. Second, the only
information required a priori is the period $\tau$ of the target UPO,
rather than a detailed knowledge of the profile of 
the orbit, or even any knowledge of the form of the original ODEs, 
which may be useful in experimental setups. The method has been implemented successfully
in a variety of laboratory experiments on electronic
\cite{PT93,GSCS94}, laser \cite{BDG94}, plasma
\cite{PBA96,FSK02}, 
and chemical~\cite{SBFHLM93,LFS95}
systems, as well as in pattern-forming systems~\cite{BS96b, MS04,
  LYH96, PS06}; more examples can be found in a
recent review by Pyragus~\cite{Pyr06}.

A paper of Nakajima~\cite{Nak97} gave a supposed restriction on the method of
Pyragus. It was believed that if a UPO in a system with no feedback had an
odd number of real Floquet multipliers greater than unity, then there
was no choice of the feedback gain matrix $K$ for which the   
method of Pyragus could be used to stabilize the UPO.
However, a recent paper of Fiedler~\etal\cite{Fie06} gives a counterexample to
this restriction. They add Pyragus-type feedback to the normal form of a subcritical Hopf
bifurcation and show that the subcritical
periodic orbit can be stabilized for some values of the feedback gain
matrix. The Hopf normal form is two-dimensional, so the subcritical orbit
has exactly one unstable Floquet multiplier. The mechanism for stabilizing the orbit is through a
transcritical bifurcation with a stable delay-induced periodic orbit.
Just~\etal\cite{Jus07} investigate a series of bifurcations in this
system, which has the attractive feature that,
despite the presence of the delay terms, much of the analysis can be
carried out analytically. 

The subcritical Hopf bifurcation of a stable equilibrium is a generic
mechanism for creating UPOs with an odd number of unstable Floquet
multipliers. 
Such bifurcations occur in a number of physical systems, such as the
Belousov--Zhabotinsky reaction-diffusion equation~\cite{IHS97}, the Hodgkin--Huxley
model of action potentials in neurons~\cite{GW00}, and in NMR lasers~\cite{BHL89}. The reduction of
these  higher-dimensional dynamical
systems to the two-dimensional normal form of the Hopf bifurcation problem
is a standard procedure~\cite{MM76,GH83}. Moreover, if Pyragus-type
feedback delay terms were added to the model ODEs, then these
(infinite-dimensional) dynamical systems could likewise be reduced to
the standard two-dimensional normal form in a vicinity of a Hopf
bifurcation~\cite{HL93}, with the parameters of the feedback control
matrix $K$ modifying the coefficients in the normal form.
Despite this disconnect between center manifold reduction of delay
equations to Hopf normal form, and the example of Fiedler~\etal in
which the feedback delay terms are added directly to the Hopf normal
form, we find that the same stabilization mechanism of subcritical
Hopf orbits applies to both their example and to the one we present
for the Lorenz equations.

Specifically, we study a subcritical Hopf bifurcation of a stable
equilibrium in the Lorenz equations~\cite{L63,Spa82}, and show that
Pyragus-type feedback can stabilize the subcritical periodic orbit. As
in the example in~\cite{Fie06}, in the absence of feedback, the
bifurcating periodic orbit has exactly one real unstable Floquet
multiplier.  It also has one stable Floquet multiplier, and one
Floquet multiplier equal to one (corresponding to the neutral
direction along the orbit). The $3\times 3$ gain matrix multiplying the Pyragus
feedback terms can be chosen in many different ways.
We give two examples in which we choose the structure of the gain
matrix in different ways and show that they
give quite different results. 

In our first example, we choose the gain matrix in a manner
suggested by the results in~\cite{Fie06} and~\cite{Jus07}: there is no
feedback in the stable direction of the UPO, and Pyragus-type feedback in the
direction of the unstable Floquet multiplier, which is identical in
form to the feedback in~\cite{Fie06}. In this way, the problem of
choosing the nine parameters in the $3\times 3$ gain matrix is
reduced to one of making an informed choice of the two parameters
employed in~\cite{Fie06}. We find that the subcritical orbit can be
stabilized over a wide range of values of our two
bifurcation parameters: the amplitude of the feedback gain, and the
usual control parameter $\rho$ in the Lorenz equations.
We identify a codimension-two point in the Hopf normal form
example, where two Hopf bifurcations collide, and show that the same
codimension-two point can be found in the Lorenz system with this
choice of feedback,
and the bifurcation structure is qualitatively the same in the two
cases. This codimension-two point captures the stabilization mechanism
in both examples: the periodic orbits created by the two Hopf bifurcations
exchange  stability in a transcritical bifurcation. The curve of
transcritical bifurcations in our two-parameter plane emanates from
the Hopf-Hopf codimension-two point.

Our second choice of the gain matrix is a real multiple of the
identity. This is also a natural choice, but in contrast with our first example we
show that here the subcritical orbit cannot be stabilized for any parameters
close to the original Hopf bifurcation. Our two bifurcation parameters
are again the amplitude of the gain and the parameter $\rho$ in the
Lorenz equations. We give analytical results on the location of Hopf
bifurcation curves and hence deduce the stability of the periodic
orbit as it bifurcates.

This paper is organized as follows. In section~\ref{sec:fied} we
review some results from Fiedler~\etal\cite{Fie06} and Just~\etal\cite{Jus07}.
In section~\ref{sec:lor} we give our example
system of the Lorenz equations with Pyragus feedback. We give two
examples of the choice of gain matrix. We explain for the first example how
we choose the gain matrix to stabilize the subcritical Hopf orbit, and
show that the bifurcation structure of this system is the same as that
for the normal form system. For the second example, the gain matrix is
a real multiple of the identity and we show that the subcritical orbit cannot be
stabilized. Section~\ref{sec:conc} concludes.

\section{The Hopf normal form with delay}
\label{sec:fied}

In this section we recap the results of~\cite{Fie06} and
identify a particular codimension-two point in the Hopf normal form
with delay which we will later examine for the Lorenz equations
with feedback. This codimension-two point acts as an organizing center
for the bifurcations involved in the mechanism for stabilizing the
periodic orbit.

The normal form of a subcritical Hopf bifurcation with a
Pyragus-type delay term is:
\begin{equation}\label{eq:hd}\dot{z}(t)=(\lambda+i)z(t)+(1+i\gamma)|z(t)|^2
z(t)+b(z(t-\tau)-z(t))\end{equation}
with $z\in\C$, and parameters $\lambda,\gamma\in\R$. The feedback gain
$b=b_0\e^{i\beta}\in\C$, and the delay
  $\tau>0$. The linear Hopf frequency has
been normalized to unity by an appropriate scaling of time. 
We consider $\lambda$ as the primary bifurcation parameter. We
consider only $\gamma<0$; this is the case in the
Lorenz example.

For the system with no feedback (\ie $b=0$) we can write
$z=r\e^{i\theta}$ and then
\begin{align}
\dot{r}&=(\lambda+ r^2)r, \\
\dot{\theta}&=1+\gamma r^2.
\end{align}
Periodic orbits exist with amplitude $r^2=-\lambda$ if $\lambda<0$, so
$\dot\theta=1-\gamma\lambda$ and the orbits have minimal period
$T=2\pi/(1-\gamma\lambda)$. We refer to these orbits as the
\emph{Pyragus orbits}, and it is these orbits that we wish to
stabilize non-invasively by adding an appropriate feedback term (\ie
with $b\neq 0$).

\begin{figure}
\psfrag{l}{\raisebox{-0.3cm}{$\lambda$}}
\psfrag{t}{\raisebox{0.15cm}{$\tau/\pi$}}
\psfrag{u}{\scriptsize{u}}
\psfrag{s}{\scriptsize{s}}
\subfigure[$b_0=0.1$]{\epsfig{figure=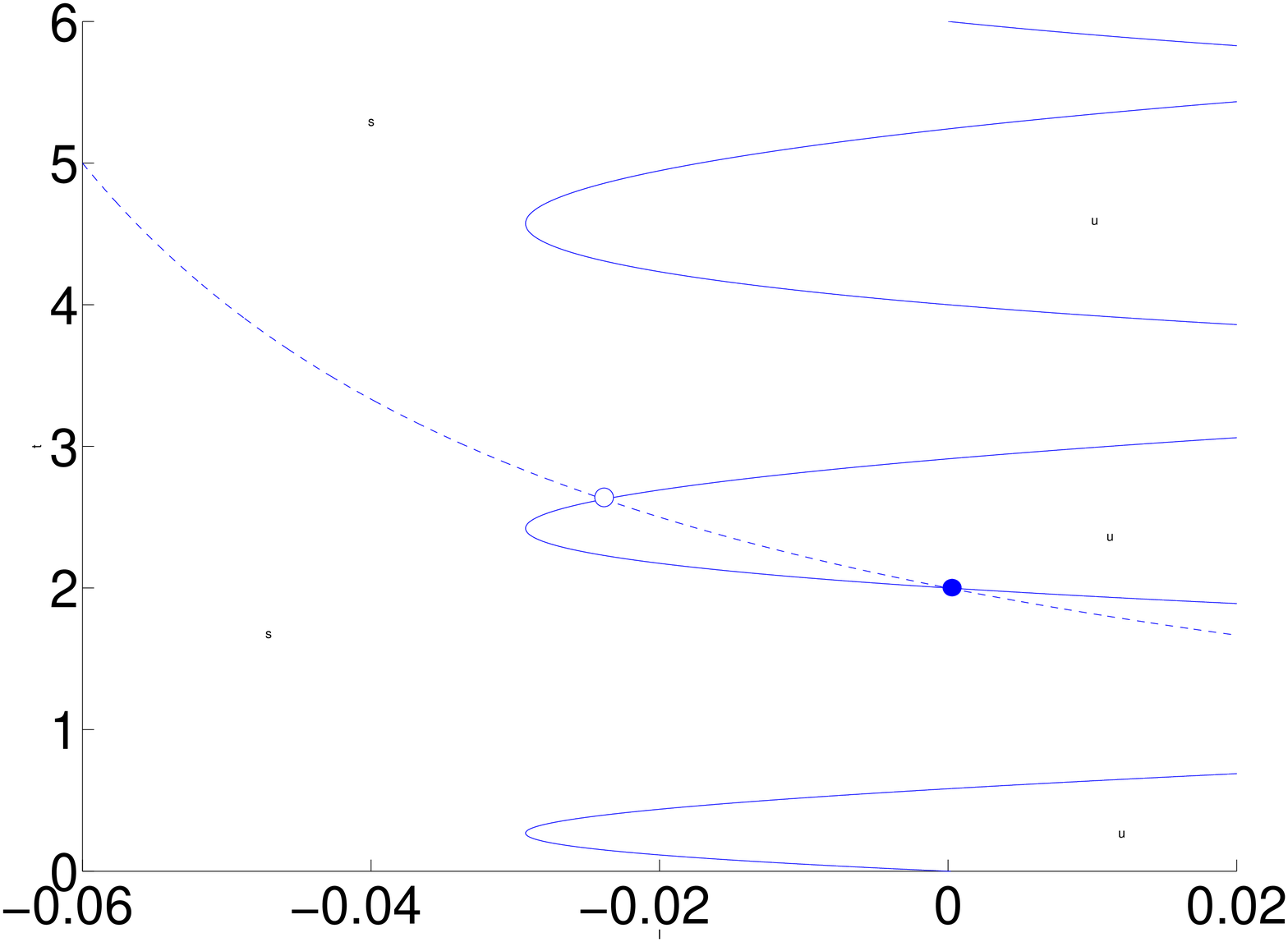,width=.32\textwidth}}
\subfigure[$b_0=0.025$]{\epsfig{figure=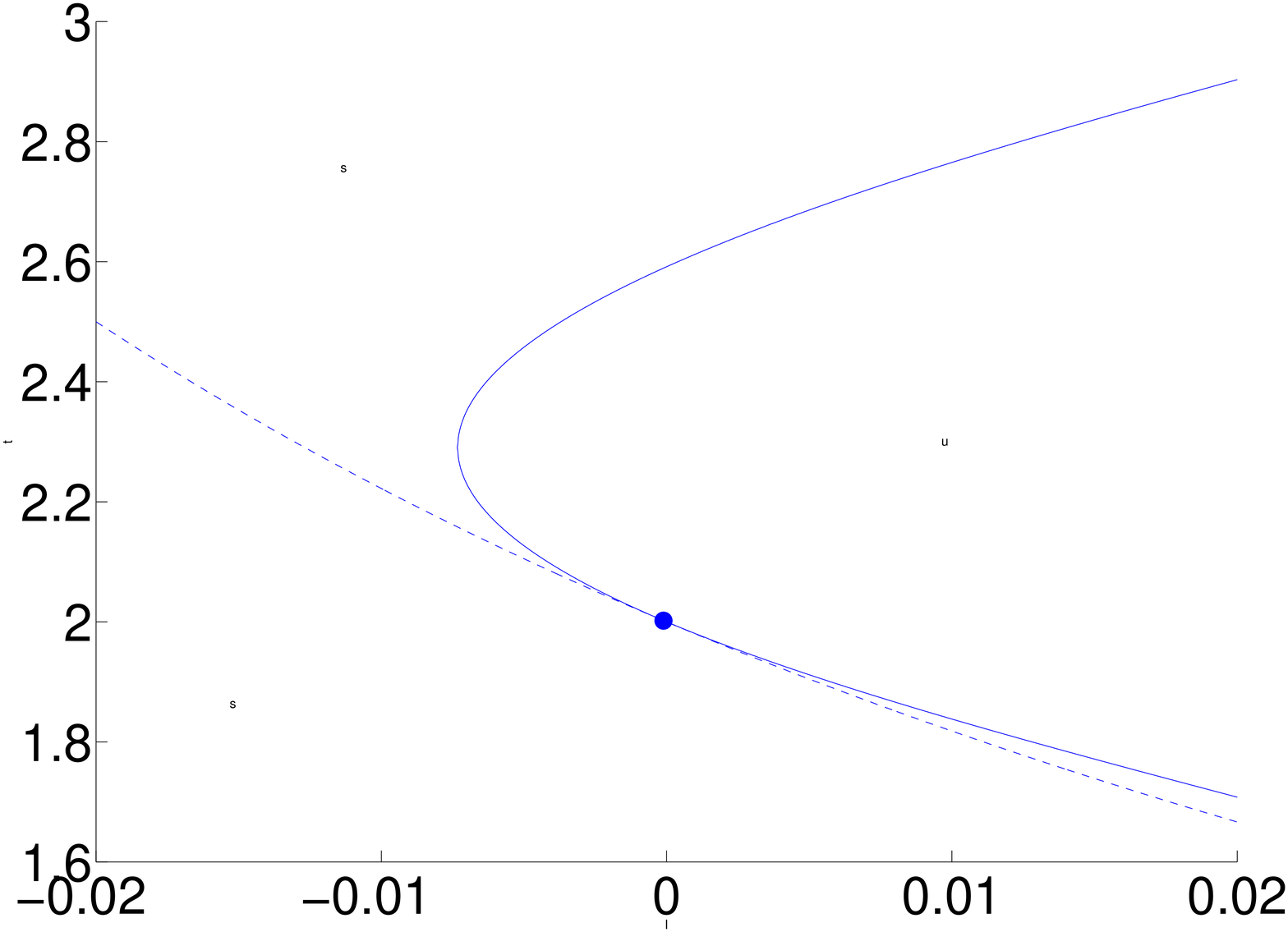,width=.32\textwidth}}
\subfigure[$b_0=0.0214$]{\epsfig{figure=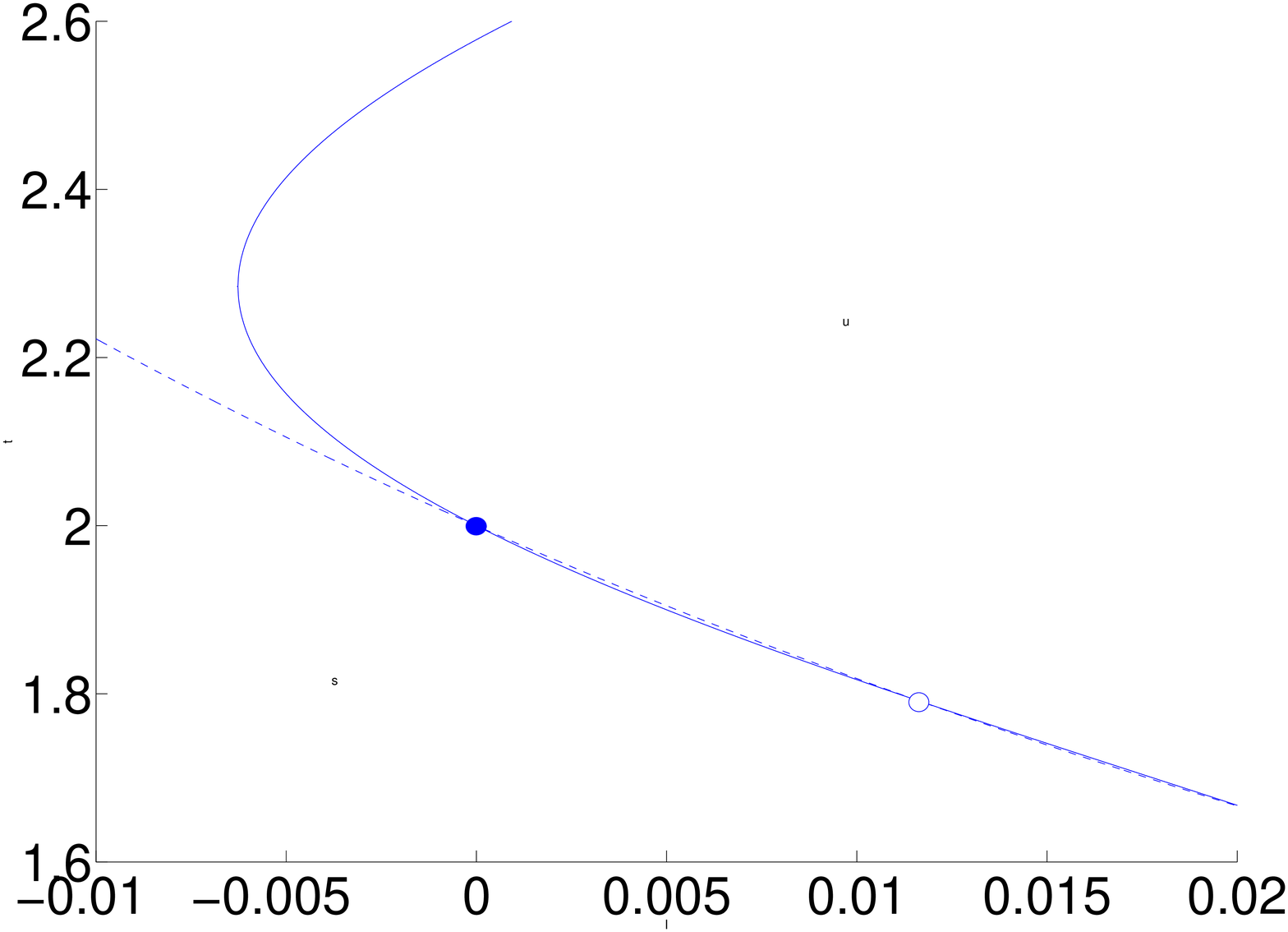,width=.32\textwidth}}
\caption{\label{fig:thtp} The figures show the curves
  $\tau=\tau_P(\lambda)$ (dashed curve) and $\tau=\tau_H(\lambda)$
  (solid curve) for three values
  of $b_0$. The remaining parameters in~\eqref{eq:hd} are $\beta=\pi/4$ and
  $\gamma=-10$. In (a), portions of four of the Hopf curves are shown, but in (b) and
  (c) we show only the curve which passes through $(\lambda,\tau)=(0,2\pi)$.
In all three cases, the two curves cross at $\lambda=0$ (shown by a
  solid dot). In (a), (with $b_0>b_0^c$), the curves cross again in
  $\lambda<0$ (shown by an empty dot), and in (c), ($b_0<b_0^c$)  the curves cross
  again in $\lambda>0$. Case (b) has $b_0=b_0^c$ and the two curves
  are tangent at $\lambda=0$. The origin is stable (unstable) in those
  regions marked by an s (u).}
\end{figure}

Following~\cite{Fie06} and~\cite{Jus07}, we define the \emph{Pyragus curve}
$\tau=\tau_P(\lambda)$ in $\lambda$-$\tau$ space, along which the
feedback vanishes on the Pyragus orbits: 
\begin{equation}\label{eq:taup}
\tau_P(\lambda)=\frac{2\pi}{1-\gamma\lambda}.
\end{equation}
We plot this curve in $\lambda$-$\tau$ space in figure~\ref{fig:thtp},
along with curves of Hopf bifurcations from the zero solution. In
later sections, we set $\tau=\tau_P(\lambda)$, as our main purpose is
the non-invasive stabilization of the Pyragus orbits.

The zero solution of~\eqref{eq:hd} undergoes Hopf bifurcations  when the
characteristic equation has purely imaginary solutions. Setting
$z(t)=\e^{\eta t}$ in~\eqref{eq:hd} and linearizing we find:
\[\eta=\lambda+i+b(\e^{\eta\tau}-1).\]
Writing $\eta=i\omega$ and separating into real and imaginary parts
gives
\begin{align}
0&=\lambda+b_0[\cos(\beta-\omega\tau)-\cos\beta], \label{eq:hf1} \\
\omega -1&=b_0[\sin(\beta-\omega\tau)-\sin\beta]. \label{eq:hf2}
\end{align}
These equations define the Hopf curves
$\tau=\tau_H(\lambda)$, in $\lambda$-$\tau$ space, parameterized by
the linear frequency
$\omega$ associated with the bifurcating periodic orbit. There are
multiple branches to
this curve, which we show in figure~\ref{fig:thtp}(a), but we concentrate on the one 
which intersects the curve $\tau=\tau_P(\lambda)$ at $(\lambda,\tau)=(0,2\pi)$.
The solution of the characteristic equation at
$\lambda=0$, $\tau=2\pi$ has $\omega=1$ and corresponds to
the Hopf bifurcation to the Pyragus orbit. 
 
Figure~\ref{fig:thtp} shows the possible
configurations of the curves $\tau=\tau_P(\lambda)$ and
$\tau=\tau_H(\lambda)$ as the parameter $b_0$ is varied.
The curves typically cross in
two places: at $\lambda=0$, and at a second location depending
on $b_0$. At $b_0=b_0^c$, the two curves are tangent at $\lambda=0$
and only intersect once. Just~\etal\cite{Jus07} show that
\[b_0^c=\frac{-1}{2\pi(\gamma\sin\beta+\cos\beta)}.\]
For simplicity, we assume $b_0^c>0$, so we must have
$\gamma\sin\beta+\cos\beta<0$. 

We define a curve of Hopf bifurcations $b_0=b_0^{\mathrm{Hopf}}(\lambda)$ in $\lambda$-$b_0$
space by the location of the
second intersection of $\tau_P(\lambda)$ and $\tau_H(\lambda)$. This
is a Hopf bifurcation to a \emph{delay-induced periodic orbit}, that
is, a periodic orbit arising from the addition of 
the delay terms; one for which the feedback does not vanish.

We adopt the convention that a Hopf bifurcation of a stable
equilibrium is called `supercritical' (`subcritical') if the resulting
periodic orbit bifurcates into the parameter regime where it coexists
with the unstable (stable) equilibrium. Such a supercritical
bifurcation generically produces a stable periodic orbit~\cite{GH83},
while the subcritical case produces an unstable periodic orbit. In the
absence of feedback, the bifurcating orbit is subcritical and
unstable. The mechanism for stabilization involves the additional
delay-induced Hopf bifurcation at $b_0^{\mathrm{Hopf}}(\lambda)$. This
bifurcation can change the trivial equilibrium from being stable to
being unstable. Consequently, the Pyragus orbit may then co-exist with
an unstable periodic orbit. As the Hopf bifurcation at
$b_0^{\mathrm{Hopf}}(\lambda)$ passes through $\lambda=0$, the
original Hopf bifurcation to the Pyragus orbit at $\lambda=0$ changes
from a subcritical one to a supercritical one. This is all done
without otherwise altering the form of the Pyragus orbit.

The Hopf bifurcation of the zero solution to the Pyragus orbit at
$(\lambda,\tau)=(0,2\pi)$ changes from subcritical to supercritical as
described above as $b_0$ is increased through $b_0^c$, since for
$b_0>b_0^c$, the curve $\tau_P(\lambda)$ lies
`inside' $\tau_H(\lambda)$. In this sense, $b_0^c$ is the
smallest value of the feedback gain for which the Pyragus orbit is
stabilized immediately after the bifurcation point. The minimum
positive $b_0^c$ can be selected by choosing $\beta$ such that $\gamma=\tan\beta$.

\subsection{A codimension-two bifurcation point}
\label{sec:cod2}

We now review some of the details of the bifurcation structure of the
system~\eqref{eq:hd} which are described in
Just~\etal\cite{Jus07}, and identify the codimension-two point we
examine in the Lorenz system. We consider $\lambda$ and $b_0$ as two
bifurcation parameters, and fix $\tau=\tau_P(\lambda)$.

The mechanism by which the Pyragus orbit is
stabilized is through a transcritical bifurcation with a delay-induced
periodic orbit. As shown
in Just~\etal\cite{Jus07}, the transcritical bifurcations occur when
\[\tau=\frac{-1}{b_0(\cos\beta+\gamma\sin\beta)},\]
or, in $\lambda$-$b_0$ space, since $\tau=\tau_P(\lambda)$, when 
\[\lambda=\frac{1}{\gamma}(1+2\pi b_0(\cos\beta+\gamma\sin\beta))=
\frac{1}{\gamma}\left(1- \frac{b_0}{b_0^c}\right).\] 
This line of transcritical bifurcations collides in $\lambda$-$b_0$ space with the two curves
of Hopf bifurcations $\lambda=0$ and
$b_0=b_0^{\mathrm{Hopf}}(\lambda)$ at $(\lambda, b_0)=(0,b_0^c)$, at a
double-Hopf codimension-two point. 
\begin{figure}
\psfrag{1}{$1$}
\psfrag{2}{$2$}
\psfrag{3}{$3$}
\psfrag{4}{$4$}
\psfrag{5}{$5$}
\psfrag{c}{$C$}
\psfrag{bh}{$b_0=b_0^{\mathrm{Hopf}}(\lambda)$}
\psfrag{l}{$\lambda$}
\psfrag{b}{$b_0$}
\psfrag{tc}{}
\psfrag{p}{Pyragus orbit}
\psfrag{np}{Delay-induced orbit}
\psfrag{z}{$z=0$}
\begin{center}
\subfigure[\label{fig:bif_dig}]{\epsfig{file=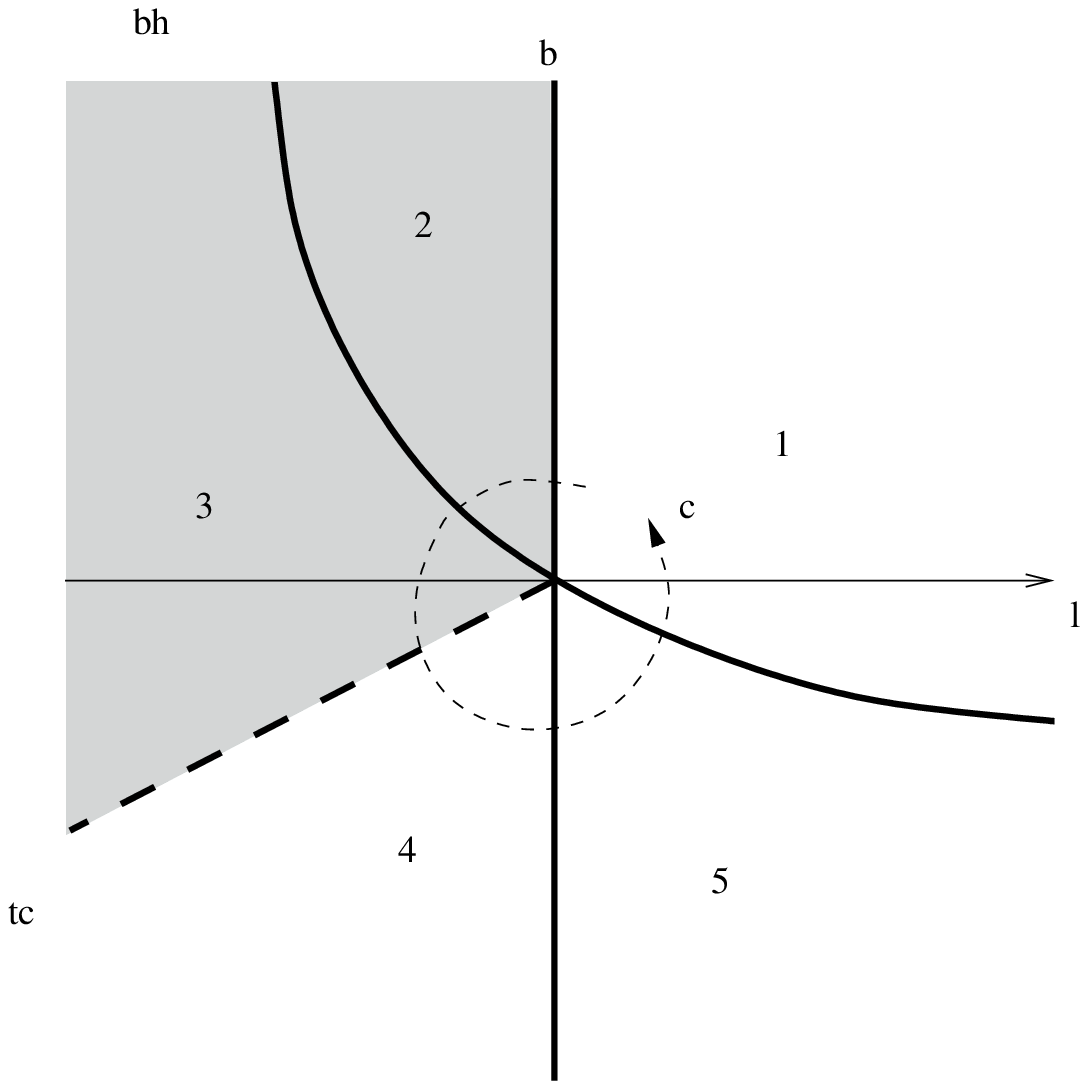,
      width=0.45\textwidth}} \quad
\raisebox{1cm}{\subfigure[\label{fig:bif_pos}]{{\epsfig{file=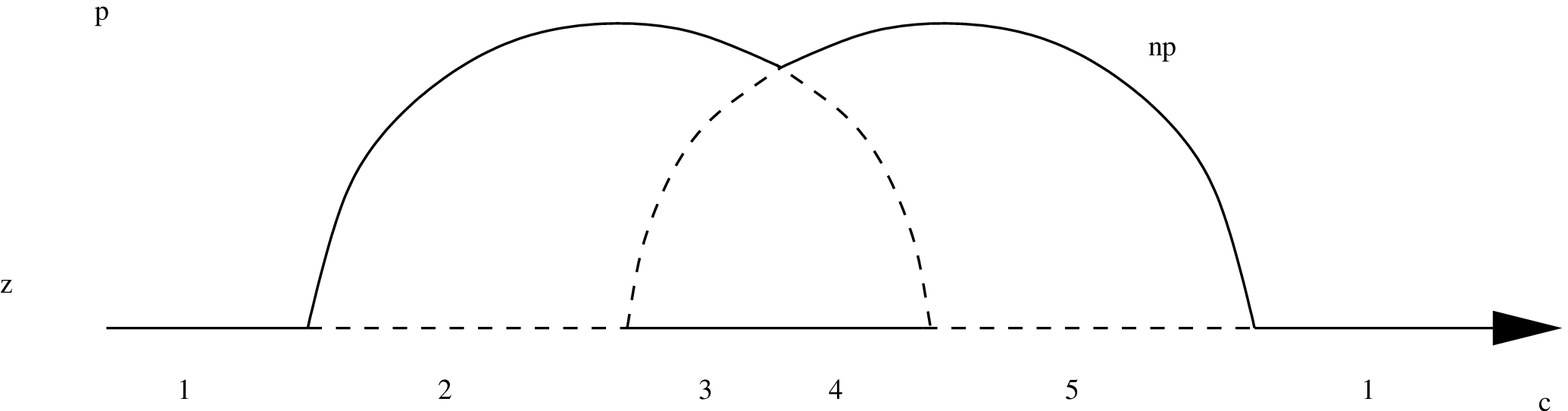, width=0.45\textwidth}}}}
\caption{\label{fig:bifs} In (a), the two Hopf bifurcation curves $\lambda=0$ and
  $b_0=b_0^{\mathrm{Hopf}}(\lambda)$ (solid bold lines), and the transcritical (TC)
  bifurcation curve (dashed bold line) divide the $\lambda$-$b_0$ plane into five
  regions. The curves intersect at $(\lambda,b_0)=(0,b_0^c)$. The
  Pyragus orbit is stable in the shaded region. (b) shows
  a schematic representation of the solutions as a path $C$ is
  traversed anticlockwise around the origin. Solid lines represent
  stable solutions and dashed lines represent unstable solutions.}
\end{center}
\end{figure}
In figure~\ref{fig:bifs} we
sketch the bifurcation structure around this point in $\lambda$-$b_0$
space. From this figure we can see that in order for the Pyragus
orbit to bifurcate stably (\ie supercritically) at $\lambda=0$, we must have $b_0>b_0^c$.

\section{The Lorenz Equations with time-delayed feedback}
\label{sec:lor}

We now use the Lorenz equations as an example system to demonstrate
that the feedback described above can also stabilize orbits arising in
a subcritical Hopf bifurcation in a higher-dimensional system of
differential equations. We give two examples of a choice of feedback
gain matrix. The first choice is informed by the results given above,
and for the second choice we set the gain matrix equal to a real
multiple of the identity.
In the first example, we further locate the
codimension-two point described in section~\ref{sec:cod2}, in the Lorenz system with
feedback, and show that the bifurcation structure is the same as in
the normal form case.

\begin{figure}
\psfrag{rho}{\raisebox{-0.15cm}{$\rho$}}
\psfrag{T}{$T$}
\psfrag{amp}{\hspace{-0.5cm}\small{amplitude}}
\begin{center}
\epsfig{file=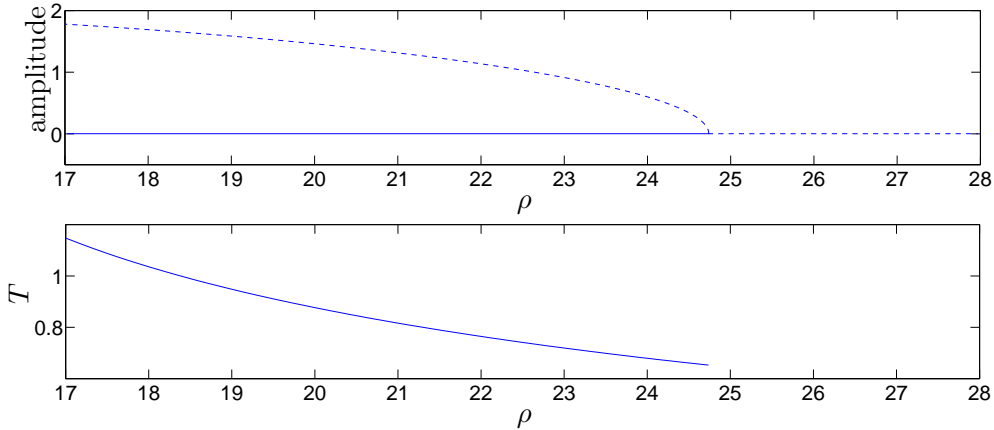, width=0.8\textwidth}
\end{center}
\caption{\label{fig:sub}The top figure is a bifurcation diagram of the
subcritical Hopf bifurcation in~\eqref{eq:lor_scaled}, showing the fixed point at zero and the
bifurcating branch of unstable periodic orbits. Solid lines indicate stable solutions and
dashed lines indicate unstable solutions. The lower figure shows the
period, $T$, of the bifurcating periodic orbit, as a function of $\rho$.}
\end{figure}

The Lorenz equations~\cite{L63,Spa82} are most often written in the following form:
\begin{align*}
\dot x & =\sigma(y-x), \\
\dot y & =\rho x - y - xz, \\
\dot z & =-\alpha z +xy,
\end{align*}
for real parameters $\sigma$, $\alpha$ and $\rho$. Lorenz and most
other authors studied the parameter regime $\sigma=10$, $\alpha=8/3$,
$\rho>0$, and we continue in the same manner. Taking $\rho$ as the
primary bifurcation parameter, the zero solution is stable for
$\rho<1$ and loses stability in a supercritical pitchfork bifurcation at
$\rho=1$. Two further equilibria are created at
\[\vec{x}_{\pm}=(\pm\sqrt{(\alpha(\rho-1))},\pm\sqrt{(\alpha(\rho-1))},\rho-1).\]
As $\rho$ is increased further, these equilibria each undergo a subcritical
Hopf bifurcation at
$\rho_h=\sigma(\sigma+\alpha+3)/(\sigma-\alpha-1)\approx 24.74$
(see~\cite{GH83} and~\cite{MM76} for further
details). It is this bifurcation that we study in the following,
 so we shift coordinates to be centered around $\vec{x}_+$ and
rescale to obtain:
\begin{align}
\dot u & =\sigma(v-u), \nonumber \\
\dot v & =u - v - (\rho-1)w-(\rho-1) uw, \label{eq:lor_scaled} \\
\dot w & = \alpha(u+v-w+uv). \nonumber
\end{align}
 Figure~\ref{fig:sub} shows a bifurcation diagram of the subcritical
 bifurcation of the zero solution of~\eqref{eq:lor_scaled}, and also the
 period $T$ of the bifurcating orbits, which we use to determine the
 delay time $\tau_P(\rho)$ in the controlled system. The periodic
 orbit exists for $13.926<\rho<\rho_h$; at the lower boundary it
 collides with a fixed point in a homoclinic bifurcation.

\subsection{Adding time-delayed feedback}

We now add Pyragus-type feedback to the Lorenz equations. We write
\begin{equation} \label{eq:lordel}
\begin{pmatrix} \dot u \\ \dot v \\ \dot w \end{pmatrix}= J(\rho)
\begin{pmatrix}  u \\  v \\ w \end{pmatrix} +
N(u,v,w)+\Gamma\begin{pmatrix}u_{\tau}-u \\v_{\tau}-  v \\w_{\tau}- w
\end{pmatrix},  
\end{equation}
where
\begin{equation}
J(\rho)=\begin{pmatrix} -\sigma & \sigma & 0 \\ 1 & -1 & -(\rho-1) \\ \alpha
& \alpha & -\alpha \end{pmatrix},\qquad N(u,v,w)=\begin{pmatrix}  0 \\
-(\rho-1)uw \\ \alpha u v \end{pmatrix},
\end{equation}
$u_{\tau}=u(t-\tau)$, \etc and $\Gamma$ is a $3\times 3$ real feedback
gain matrix, to be determined. In this section we use the results of~\cite{Fie06} and~\cite{Jus07} to inform our
choice of the control matrix $\Gamma$. In general, $\Gamma$ would
contain nine independent parameters, but the method we describe
reduces this to only two. In section~\ref{sec:id_feed} we describe the
dynamics when $\Gamma$ is a real multiple of the identity.
Note that if this system were reduced
to normal form around the Hopf bifurcation point, the resulting
equations would not be the same as~\eqref{eq:hd}. That is, there would
be no delay terms; the delay terms here would only have the affect of
altering the parameters in the usual Hopf normal form. See~\cite{HL93} for more details.

At the bifurcation point ($\rho=\rho_h$), $J$ has one real negative
eigenvalue (which we denote by $-\lambda_h=-\alpha-\sigma-1\approx -11.7$), and a pair of purely imaginary
eigenvalues ($\pm i\omega_h$,
$\omega_h=\sqrt{(2\alpha\sigma(\sigma+1))/(\sigma-\alpha-1)}\approx 9.62$). The center manifold of
the original problem with no feedback is therefore  
two-dimensional, and the eigenvectors of $J$ can be found explicitly
(see~\cite{MM76}). Close to the bifurcation point, the subcritical
orbit will lie in a two-dimensional manifold which is close to
the center subspace at the bifurcation point. We therefore choose
\[\Gamma=EGE^{-1},\] where $E$ is the matrix of eigenvectors which puts
$J(\rho_h)=J_h$ in Jordan normal form, that is
\begin{equation}
E^{-1}J_hE=\begin{pmatrix} -\lambda_h & 0 & 0 \\ 0 & 0 & -\omega_h \\
0 & \omega_h & 0 \end{pmatrix},
\end{equation}
and
\begin{equation}
G=\begin{pmatrix} 0 & 0 & 0 \\ 0 & b_0\cos\beta & -b_0\sin\beta \\
0 & b_0\sin\beta & b_0\cos\beta \end{pmatrix}.
\end{equation}
There is then no feedback in the stable direction, and Pyragus-type
feedback in the directions tangent to the center manifold.

In the following numerical results, we set $\beta=\pi/4$, as in
Fiedler~\emph{et al}., and vary $b_0$.

\subsection{Numerical results}

We use the continuation package DDE-BIFTOOL~\cite{ddebiftool} to analyze the
delay-differential equation~\eqref{eq:lordel}. The primary bifurcation
parameter is $\rho$, with the subcritical Hopf bifurcation for the
system without feedback occurring at $\rho=\rho_h\approx
24.74$. Recall we have set $\sigma=10$, $\alpha=8/3$ and $\beta=\pi/4$.

First, we locate Hopf bifurcations of the trivial solution in the $\rho$-$\tau$ plane, for
various values of $b_0$. 
\begin{figure}
\psfrag{tau}{$\tau$}
\psfrag{rho}{\raisebox{-0.2cm}{$\rho$}}
\psfrag{tp}{}
\psfrag{th}{}
\psfrag{u}{\scriptsize{u}}
\psfrag{s}{\scriptsize{s}}
\begin{center}
\subfigure[$b_0=1.2$]{\epsfig{file=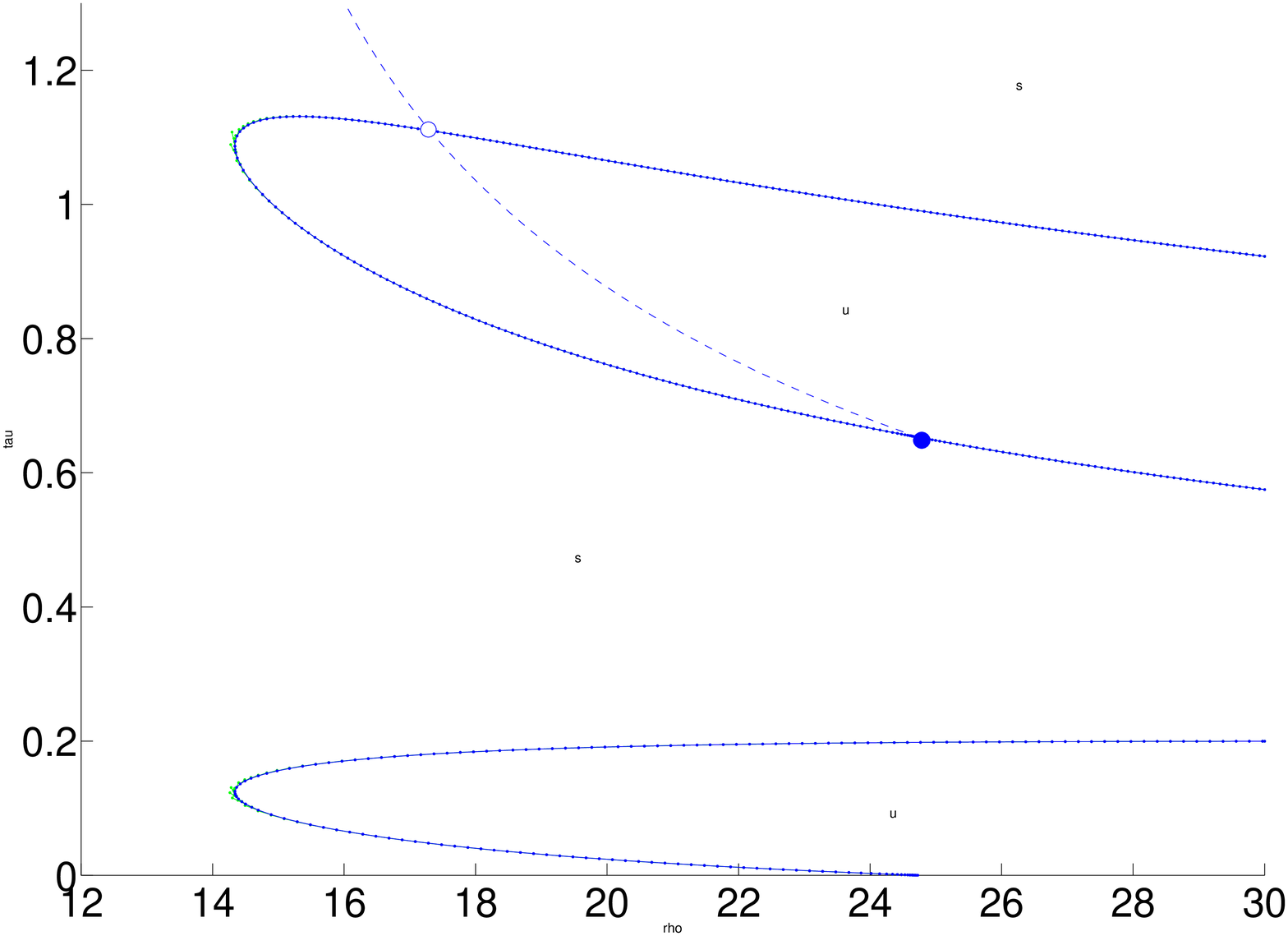, width=0.45\textwidth}}
\subfigure[$b_0=0.1$]{\epsfig{file=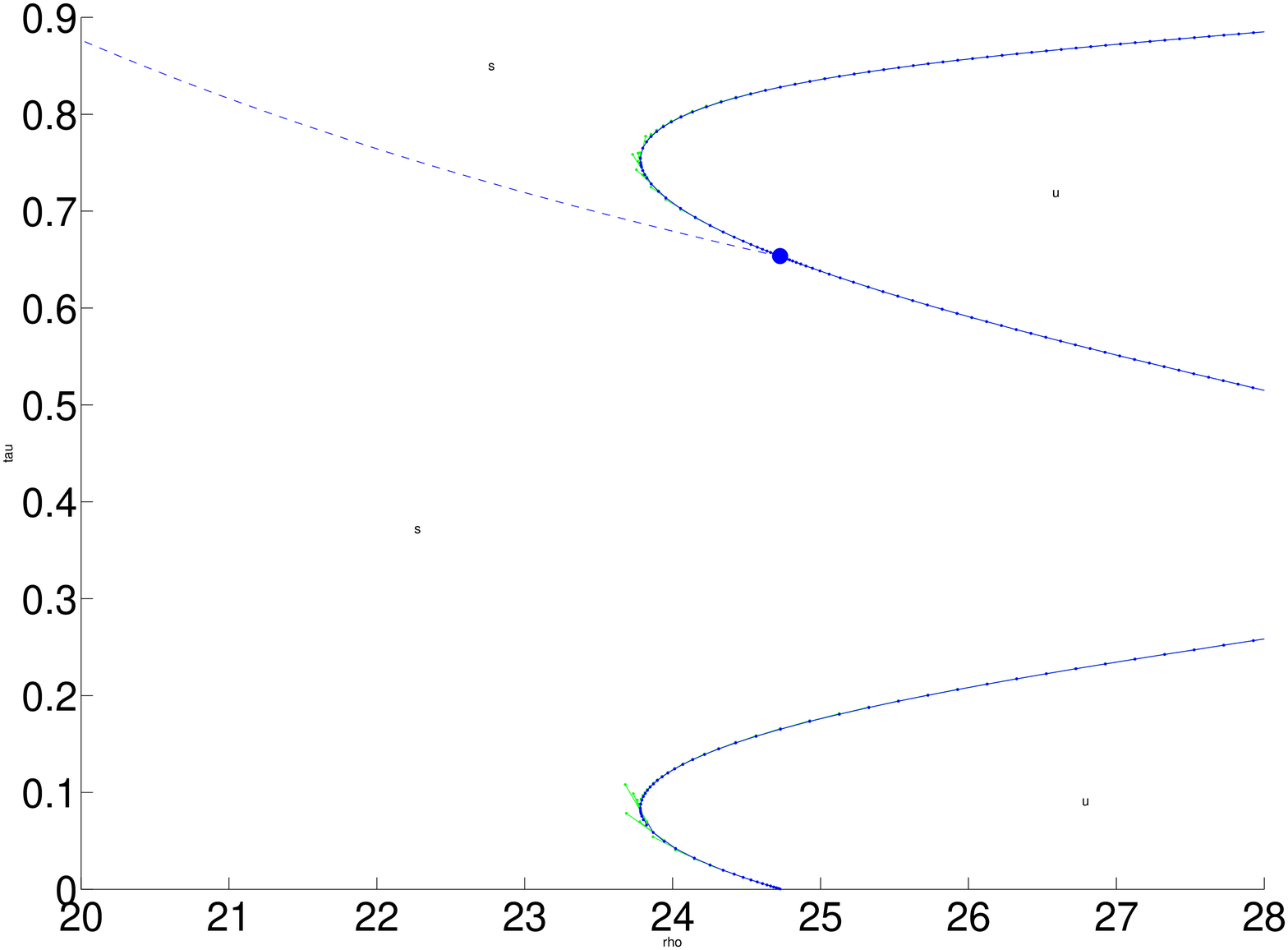, width=0.45\textwidth}}
\end{center}
\caption{\label{fig:ThTp}The figure shows the curves
  $\tau=\tau_H(\rho)$ (solid lines) and $\tau=\tau_P(\rho)$ (dashed lines) for $b_0=1.2>b_0^c$ and
  $b_0=0.1<b_0^c$. Remaining parameters are $\sigma=10$, $\alpha=8/3$ and $\beta=\pi/4$.
Compare with figure~\ref{fig:thtp}. The Hopf
  bifurcation at $\rho=\rho_h\approx 24.74$ is shown with a solid
  dot. In (a) the curves additionally cross in $\rho<\rho_h$ (shown by
  an empty dot). The origin is stable (unstable) in those regions
  marked with an s (u).
  These figures were produced using DDE-BIFTOOL~\cite{ddebiftool}.}
\end{figure}
Figure~\ref{fig:ThTp} shows curves of Hopf
bifurcations $\tau=\tau_H(\rho)$ for $b_0=1.2$ and $b_0=0.1$. We also
plot the curve $\tau=\tau_P(\rho)$, given by the period of the
bifurcating subcritical orbits (see 
figure~\ref{fig:sub}).  Figure~\ref{fig:ThTp} is qualitatively similar to
figure~\ref{fig:thtp} (the corresponding figure for the normal form case). 
For $b_0=1.2$, $\tau_P(\rho)$ lies inside $\tau_H(\rho)$, and so we
expect, by analogy with the normal form case, that
choosing $\tau=\tau_P(\rho)$ will stabilize the Pyragus
orbit. For $b_0=0.1$, $\tau_P(\rho)$ lies outside
$\tau_H(\rho)$, and so the feedback cannot stabilize the Pyragus
orbit near onset, since it bifurcates subcritically (\ie it coexists
with the stable equilibrium from which it bifurcates).
The codimension-two point occurs at some value of $b_0$
that is the boundary between these cases. 

We use DDE-BIFTOOL to locate this codimension-two point. As in the
normal form case, we set $\tau=\tau_P(\rho)$. We do not have an
analytic form for $\tau_P(\rho)$, so we
numerically estimate $\tau_P(\rho)$ in the following way.
For $\rho<\rho_h$ we set $\tau_P(\rho)$ equal to the period of the
bifurcating period orbits for the system with no feedback (see
figure~\ref{fig:sub}). We want to continue $\tau_P(\rho)$ into
$\rho>\rho_h$, so we can complete both sides of the bifurcation
diagram, so here we set $\tau_P(\rho)=\tau_h/(1-B(\rho-\rho_h))$,
where $B=-0.0528$, and $\tau_h=\tau_P(\rho_h)\approx 0.6528$. This choice of $B$ ensures that $\tau_P(\rho)$ is
continuous and has continuous first derivative at $\rho=\rho_h$.

\begin{figure}
\psfrag{rho}{\raisebox{-0.2cm}{$\rho$}}
\psfrag{rhoh}{\raisebox{-0.1cm}{$\rho_h$}}
\psfrag{b0}{\raisebox{0.1cm}{\hspace{-0.1cm}{$b_0$}}}
\psfrag{th}{\raisebox{-0.2cm}{$\theta$}}
\psfrag{a}{\hspace{-0.7cm}\raisebox{0.1cm}{amplitude}}
\psfrag{p}{\scriptsize{Pyragus orbit}}
\psfrag{di}{\scriptsize{Delay-induced orbit}}
\begin{center}
\subfigure[]{\epsfig{file=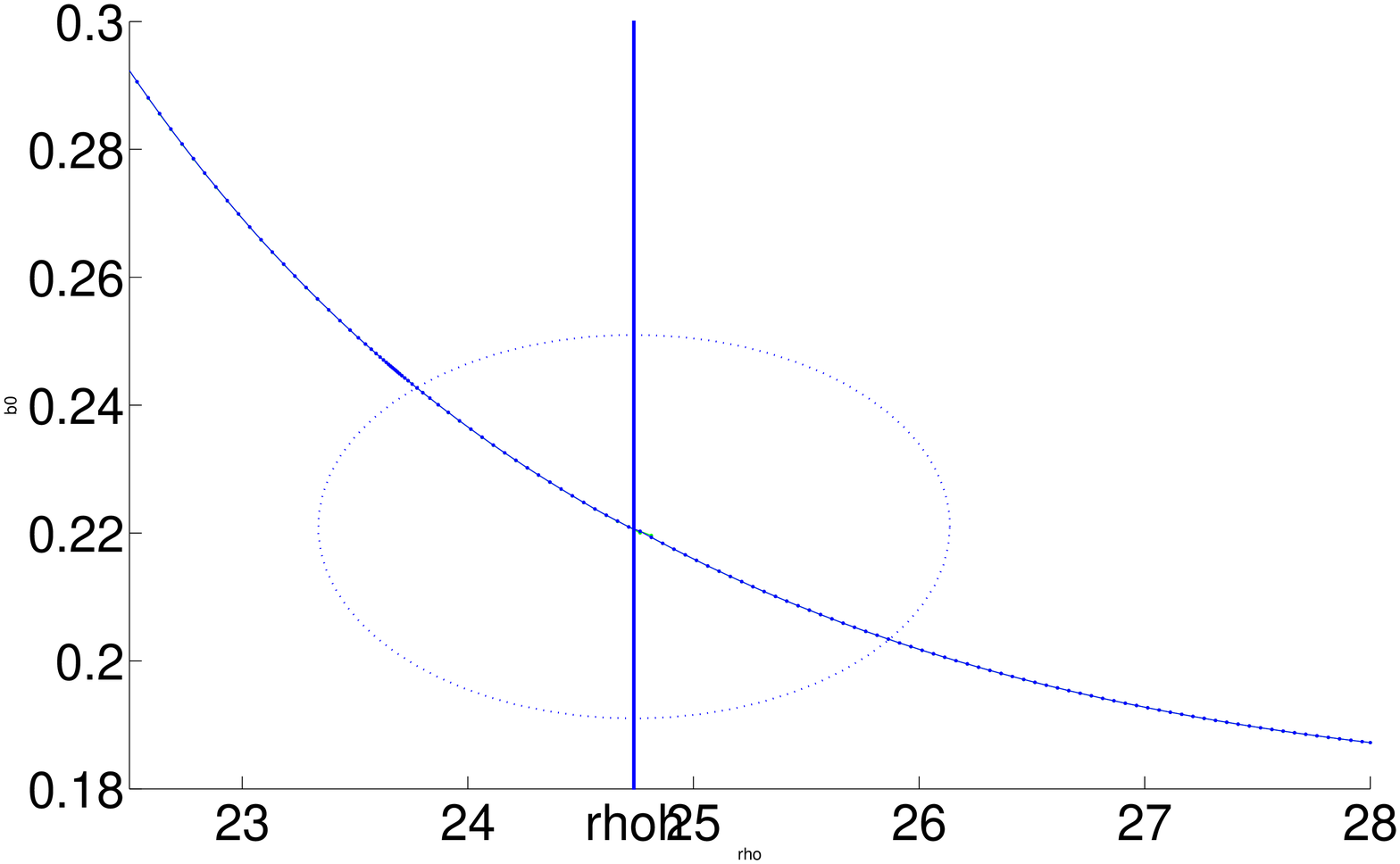, width=0.45\textwidth}}
\subfigure[]{\epsfig{file=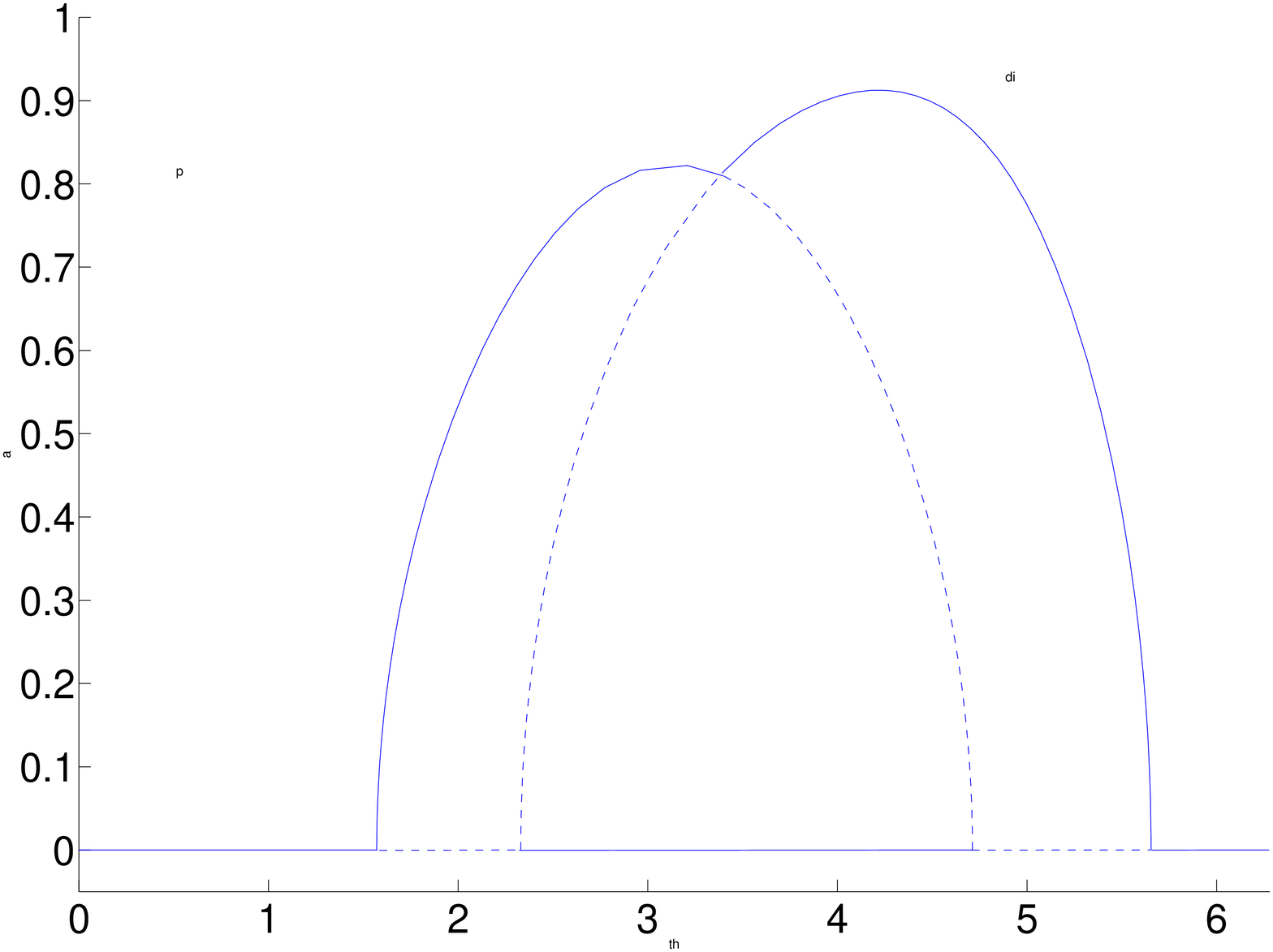, width=0.45\textwidth}}
\end{center}
\caption{\label{fig:lor_hopfs} (a) shows curves of Hopf
  bifurcations from the zero solution in the $\rho$-$b_0$ plane for
  the Lorenz system with delay. The vertical line is the
  bifurcation to the Pyragus periodic orbits, and the second
  curve is the bifurcation to the delay-induced periodic orbits. The
  curve of transcritical bifurcations of periodic orbits is not shown,
  but it can be seen in figure~\ref{fig:po_stab}. The
  dotted ellipse is the curve traversed to generate the bifurcation
  diagram in (b). Here, solid lines indicate stable
 solutions, and dashed lines indicate unstable solutions. 
Parameter values are $\sigma=10$, $\alpha=8/3$,  $\beta=\pi/4$. The
  ellipse is parameterised by $\theta$: $\rho-\rho_h=1.4\cos\theta$,
  $b_0-b_0^c=0.03\sin\theta$, and encloses the codimension-two
  point.
This figure was generated using
  DDE-BIFTOOL. Compare with figure~\ref{fig:bifs}.}
\end{figure}

With the parameter restriction $\tau=\tau_P(\rho)$, we generate curves
of Hopf bifurcations from the zero solution in the $\rho$-$b_0$ plane;
these are shown in figure~\ref{fig:lor_hopfs}. 
We can then estimate the
location of the codimension-two point, at $(\rho,b_0)=(\rho_h,b_0^c)$,
the point where the two curves of Hopf bifurcations cross. We find
$b_0^c\approx 0.221$.
We follow a path
around the codimension-two point and track the amplitude and stability
of the bifurcating periodic orbits; a bifurcation diagram of the
periodic orbits is shown in
figure~\ref{fig:lor_hopfs}. The transcritical bifurcation of periodic
orbits can clearly be seen.
Note that figure~\ref{fig:lor_hopfs} is qualitatively similar to
figure~\ref{fig:bifs}, showing that the
bifurcation structure in the normal form case, and in our Lorenz
example are the same.

With the additional feedback, the Pyragus orbits are stable for a
wide parameter range. 
\begin{figure}
\psfrag{rho}{\raisebox{-0.2cm}{$\rho$}}
\psfrag{b0}{$b_0$}
\begin{center}
\epsfig{file=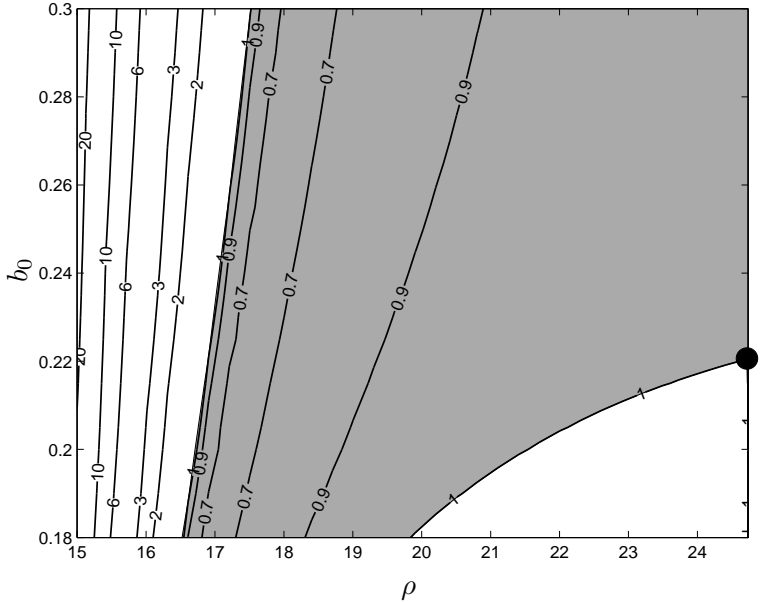, width=0.7\textwidth}
\end{center}
\caption{\label{fig:po_stab} The figure shows a contour plot of the modulus of the
  largest Floquet multipliers for the Pyragus orbit as $\rho$ and
  $b_0$ are varied. The shaded area indicates the region where the
  periodic orbit is stable. The codimension-two point is marked with a
  dot. The boundary of the stable region which emanates from this
  point is the transcritical bifurcation of periodic orbits. The left
  boundary of the stable region corresponds to a period-doubling bifurcation. This
  figure was produced using DDE-BIFTOOL.}
\end{figure}
In figure~\ref{fig:po_stab}, we show the
stability of the Pyragus orbits as $\rho$ and $b_0$ are varied. The
transcritical bifurcation can be seen as the boundary of the stable
region which terminates at the codimension-two point. The
orbits also undergo an instability at around $\rho\approx 17$. Along
this boundary of the stable region, the periodic orbits have a Floquet multiplier equal to $-1$,
and so the instability is a period-doubling bifurcation.

\begin{figure}
\psfrag{t}{\raisebox{-0.1cm}{$t$}}
\psfrag{x}{\hspace{-0.4cm}$u,v,w$}
\begin{center}
\epsfig{file=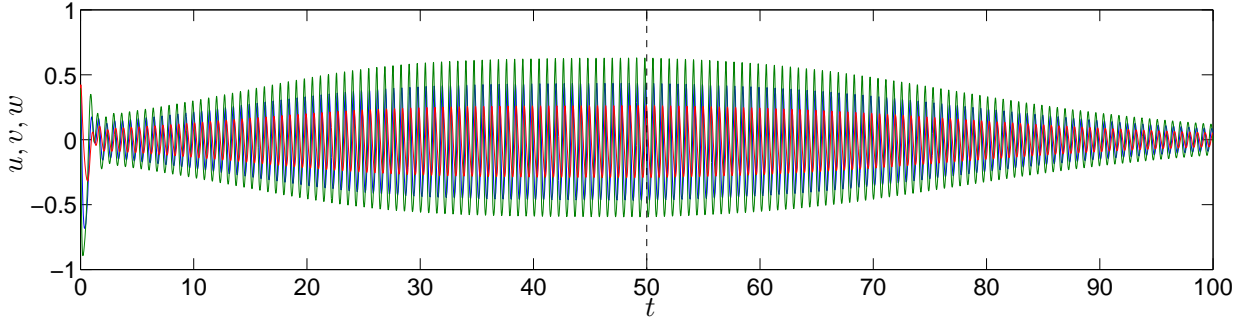, width=\textwidth}
\end{center}
\caption{\label{fig:tplot1} (Color online) The figure shows time integration of the
  Lorenz equations with feedback which is switched off at $t=50$ (indicated by a dashed vertical line).
 Parameter values are $\sigma=10$, $\alpha=8/3$,
  $\beta=\pi/4$, $\rho=23$ and  $\tau=\tau_P(\rho)=0.7191$. $b_0=1.2$
  for $t\leq 50$ and $b_0=0$ for $t>50$. The Pyragus orbit
  is initially stable, but without feedback, the trajectory decays
  back to the origin. The data was computed using the Matlab routine
  \texttt{dde23} for integrating delay-differential equations.}
\end{figure}
In figure~\ref{fig:tplot1} we show results of forward time integration
of the delay-differential equation~\eqref{eq:lordel}, at $\rho=23<\rho_h$,
with $\tau=\tau_P(\rho)=0.7191$. Initally, $b_0=1.2$, and the
Pyragus orbit is stable. The feedback is then turned off
(\ie $b_0=0$) at $t=50$, and the trajectory decays back to the zero
solution. We have used DDE-BIFTOOL to confirm the stability of these orbits.

\begin{figure}
\psfrag{t}{\raisebox{-0.2cm}{$t$}}
\psfrag{x}{\hspace{-0.4cm}$u,v,w$}
\psfrag{u}{\raisebox{-0.2cm}{$u$}}
\psfrag{w}{\raisebox{0.1cm}{$w$}}
\begin{center}
\subfigure[]{\epsfig{file=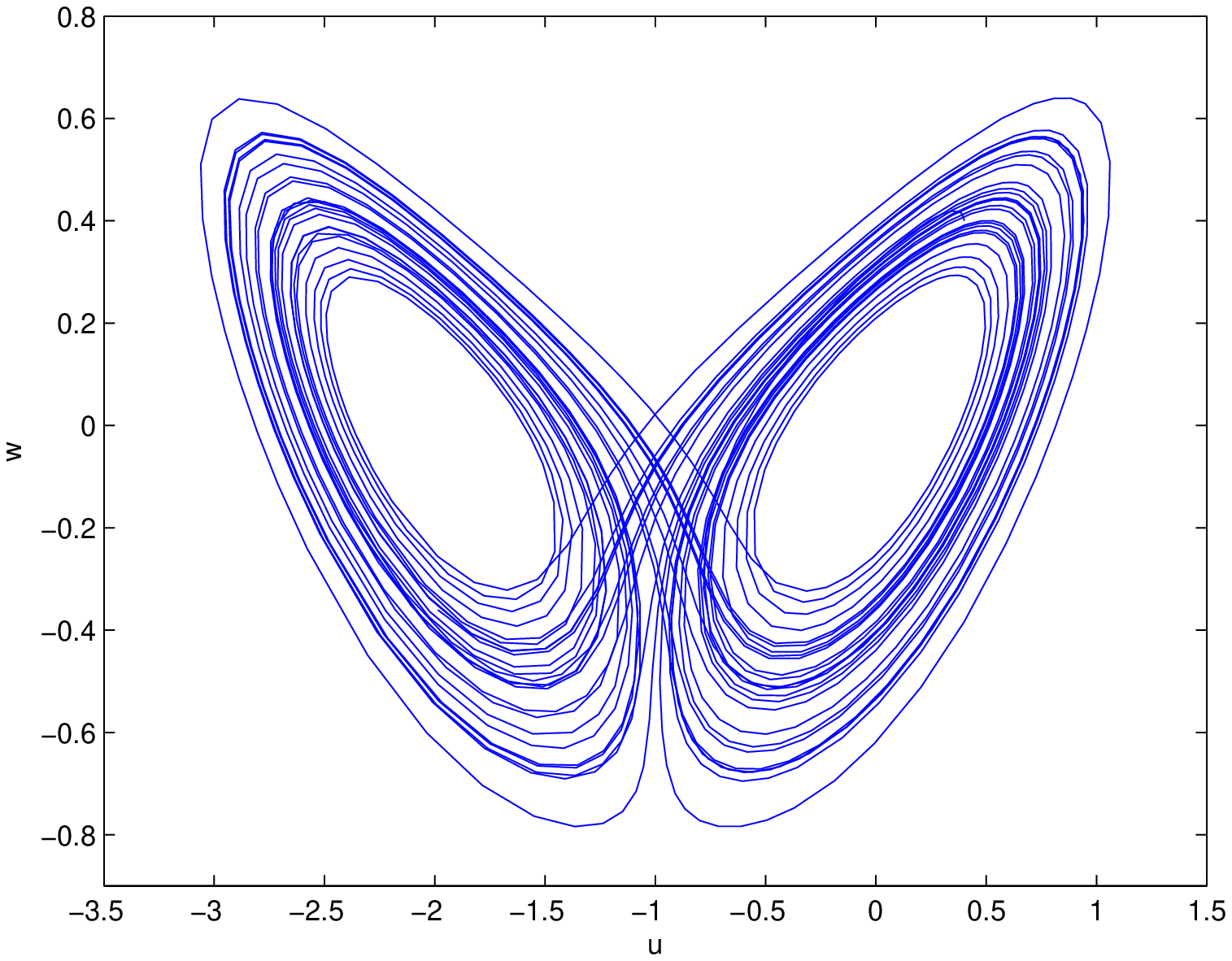, width=0.48\textwidth}}
\subfigure[]{\epsfig{file=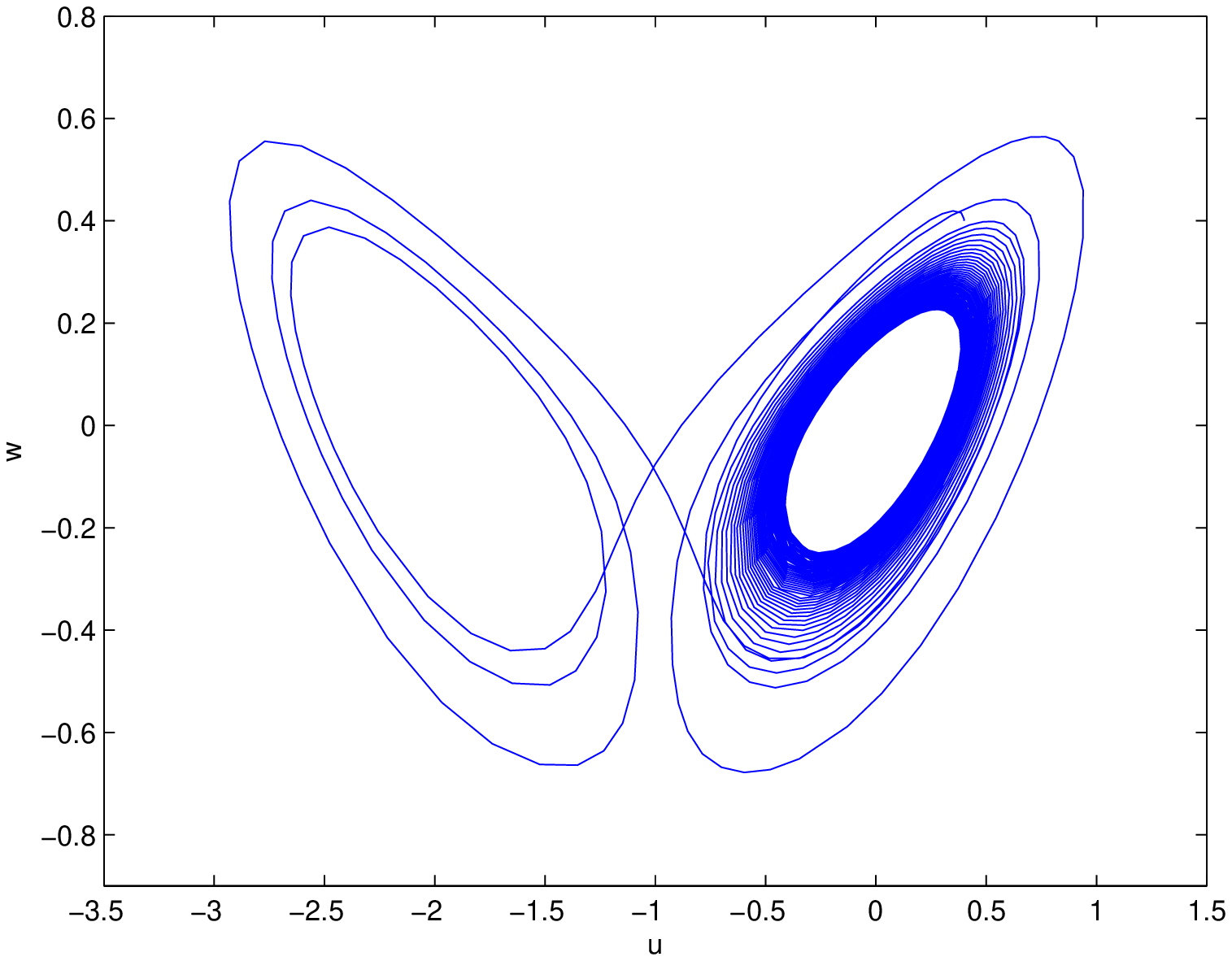, width=0.48\textwidth}}
\subfigure[]{\epsfig{file=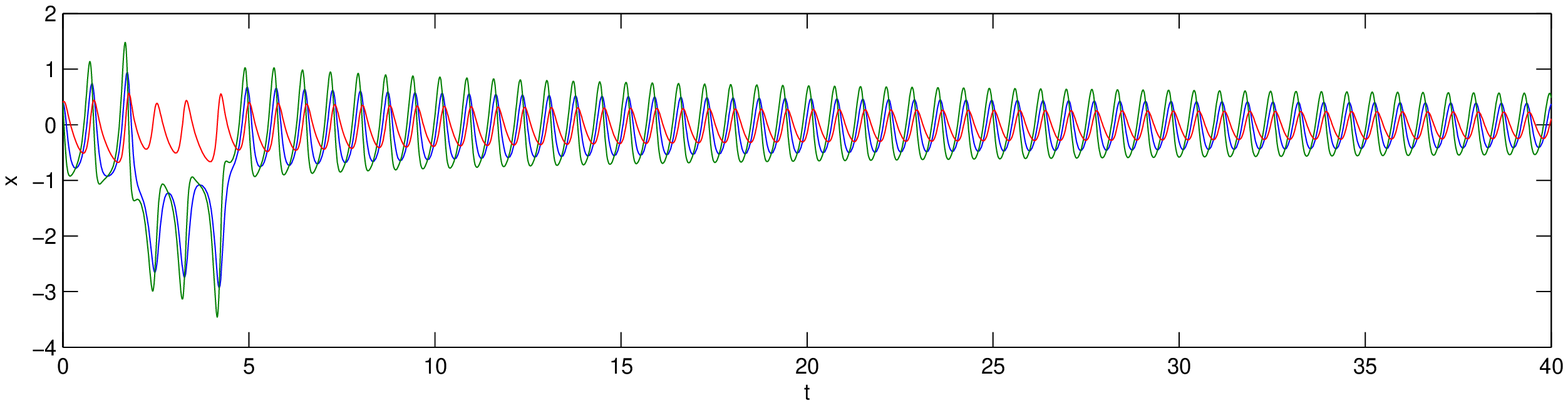, width=\textwidth}}
\end{center}
\caption{\label{fig:tplot2} (Colour online) Figures (a) and (b) show two time integrations of the
  Lorenz equations. Figure (a) has no feedback terms, and in figure
  (b) the feedback is `turned on' at $t=5$. Without feedback, the
  well-known chaotic strange attractor is stable, but with the feedback, a
  delay-induced periodic orbit becomes the stable solution. The
  stability of the periodic orbit has been confirmed using DDE-BIFTOOL. Parameters
  are $\rho=24.8388$, $b_0=0.22$, $\tau=0.6494$. The stable periodic
  orbit has a period of $0.6537$. Figure (c) shows the time plot of
  figure (b). The data was computed using the Matlab routine \texttt{dde23}.}
\end{figure}

The structure around the codimension-two point also tells us 
that the delay-induced orbits can be stable in the region $\rho>\rho_h$. For example, at
$\rho=24.8388$, $b_0=0.22$, $\tau=0.6494$, we can use DDE-BIFTOOL to show that there exists a stable
delay-induced periodic orbit with a period of
$0.6537$. Figure~\ref{fig:tplot2} shows time integration at these
parameter values, with feedback turned on at $t=5$. The
figure shows the chaotic attractor for $t<5$ and an approach to a stable
periodic orbit for $t>5$. 

\subsection{A second example}
\label{sec:id_feed}

Another natural choice for the gain matrix $\Gamma$ is a real multiple
of the identity. As a comparison to the results given above, we now
consider this case, so write
\begin{equation}\label{eq:Gid}
\Gamma=b_0\begin{pmatrix}1 & 0 & 0 \\ 0 & 1 & 0 \\ 0 & 0 & 1
\end{pmatrix},\qquad b_0\in\R.\end{equation}
For this choice of $\Gamma$ we can obtain analytical
results, and we show that the bifurcation structure is
different from our previous example, whatever the value of $b_0$. We show first that unstable periodic orbits with real positive Floquet exponents cannot be stabilised using this form of feedback. We then discuss the shape and location of the curves of Hopf bifurcation from the origin, in a similar manner to the previous section, to give a comparison of the two types of feedback.
The codimension-two point
described previously does not exist, and the Hopf 
bifurcation to the Pyragus orbit is always subcritical. That is, the
periodic orbit in $\rho<\rho_h$ always bifurcates unstably from an
equilibrium which is stable in $\rho<\rho_h$ and unstable in $\rho>\rho_h$.

Consider a Pyragus solution $\vec{u}^{\star}(t)$ of~\eqref{eq:lordel}, which is
 periodic with period $\tau$, and with $\Gamma$ as in~\eqref{eq:Gid}.
 Then if $\mu$ is a Floquet exponent of  $\vec{u}^{\star}$ (in the
 system with no feedback), the characteristic equation for  $\vec{u}^{\star}$ in 
the system with feedback is given by
\begin{equation}\label{eq:chareqpo}
(\lambda-b_0(\e^{-\lambda\tau}-1))=\mu.\end{equation}
Both here, and in the analysis which follows below, simplification is possible because $\Gamma$ is a multiple of the identity.
If $\vec{u}^{\star}$ has one Floquet exponent $\mu$ which is real and positive, then
it can be shown (see \eg~\cite{BC63}) that there always exists at least one 
solution $\lambda$ which is real and positive. Hence stabilization of 
$\vec{u}^{\star}$ cannot be achieved.

We note that at the subcritical Hopf bifurcation from the zero solution in~\eqref{eq:lor_scaled}, the orbit which bifurcates has one real stable multiplier (inherited
 from the zero solution) and one neutral multiplier, and so the remaining unstable 
multiplier must be real. Therefore the Pyragus orbit cannot be stabilized close 
to the Hopf bifurcation using this type of feedback.


In addition, we now follow the method of the previous section to find curves of Hopf bifurcations of the zero solution of~\eqref{eq:lordel} in $\rho$-$\tau$ space. This provides us with a comparison of the two types of feedback used in this and the previous section.
 We are particularly
interested in those curves which pass through
$(\rho,\tau)=(\rho_h,\tau_h)$, with Hopf frequency equal to
$\omega_h$, as this is the Hopf bifurcation to the Pyragus orbit.

Consider the linearisation of~\eqref{eq:lordel} about the origin, and
write $(u,v,w)^{T}=\vec{u}\e^{\lambda t}$. Then
\[\lambda\vec{u}\e^{\lambda t}=J(\rho)\vec{u}\e^{\lambda t}
+\Gamma\vec{u}(\e^{-\lambda\tau}-1)\e^{\lambda t},\]
and so for a non-trivial solution ($\vec{u}\neq\vec{0}$) to exist, $\lambda$
must satisfy the characteristic equation:
\begin{equation}\label{eq:chareq}
\det[(\lambda-b_0(\e^{-\lambda\tau}-1))I-J(\rho)]=0.\end{equation}
Note that we have been able to simplify this equation because in this example
$\Gamma$ is a multiple of the identity.
This tells us that $g(\lambda)\equiv(\lambda-b_0(\e^{-\lambda\tau}-1))$ are the eigenvalues of
$J(\rho)$. 

When $\rho$ is close to $\rho_h$, $J(\rho)$ has one negative
eigenvalue $-\lambda_1(\rho)$, and a complex conjugate pair we denote as
$\mu(\rho)\pm i\nu(\rho)$. Note that $\mu(\rho_h)=0$, and $\nu(\rho_h)=\omega_h$.

We find curves of Hopf bifurcations in $\rho$-$\tau$ space, by writing $\lambda=i\omega$ ($\omega\in\R$) and setting
$g(i\omega)$ equal to the eigenvalues
of $J(\rho)$, $\mu(\rho)+i\nu(\rho)$. We note that setting $g(i\omega)$ equal to the negative eigenvalue of $J(\rho)$, $-\lambda_1(\rho)$, does not produce any Hopf curves which pass through $(\rho,\tau)=(\rho_h,\tau_h)$, so we do not consider these here.
 Equating
real and imaginary parts gives
\begin{align}
-b_0(\cos(\omega\tau)-1)&=\mu(\rho), \label{eq:hfid1} \\
\omega+b_0\sin(\omega\tau)&=\nu(\rho). \label{eq:hfid2}
\end{align}
Equations~\eqref{eq:hfid1} and~\eqref{eq:hfid2} describe curves of
Hopf bifurcations in $\rho$-$\tau$ space, parameterized by the Hopf
frequency $\omega$.
\begin{figure}
\psfrag{rho}{\raisebox{-0.2cm}{$\rho$}}
\psfrag{t}{\raisebox{0.1cm}{\hspace{-0.1cm}{$\tau$}}}
\begin{center}\psfrag{nu}{\raisebox{-0.2cm}{$\nu$}}
\psfrag{lam}{\hspace{0.1cm}\raisebox{0.1cm}{$-\lambda_1$}}
\epsfig{file=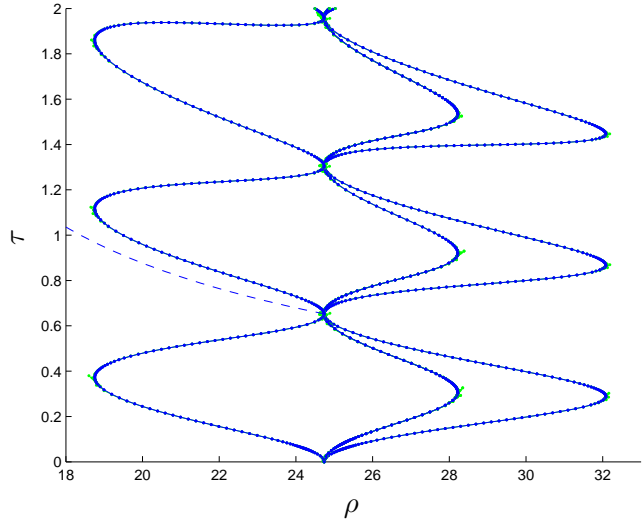, width=0.6\textwidth}
\end{center}
\caption{\label{fig:hopf_id} The figure shows curves of Hopf
  bifurcations from the zero solution of~\eqref{eq:lordel} with
  $\Gamma=b_0I$. Curves are shown for $b_0=0.1,0.05$ and $-0.1$ (from
  right to left). The zero solution is stable to the left of these
  curves and unstable to the right of these curves. The dashed curve is the 
  curve $\tau=\tau_P(\rho)$. For each value of $b_0$, the curve lies between $\rho=\rho_h$ and $\rho=\rho^{\star}(b_0)$.}
\end{figure}
Figure~\ref{fig:hopf_id} shows examples of the shape of these Hopf curves; compare with figure~\ref{fig:ThTp} showing the Hopf curves in our previous example. The zero
 solution is unstable to the right of the curves and stable to the
 left of the curves. We now explain why we expect the curves to have this shape.

We have:
\[(\nu(\rho)-\omega)^2=\mu(\rho)(2b_0-\mu(\rho)).\]
Note that  we  need $\mu(\rho)(2b_0-\mu(\rho))\geq0$ for solutions to exist. That is, if
 $b_0>0$, we need $0\leq\mu(\rho)\leq 2b_0$, and if $b_0<0$, we need
 $2b_0\leq\mu(\rho)\leq 0$.
Since $\mu(\rho)$ is a monotonically increasing function over the range of
 $\rho$ we are considering, this
 gives a connected range of $\rho$ for which Hopf bifurcations can
 occur, with boundaries at $\rho=\rho_h$ (since $\mu(\rho_h)=0$) and
 at $\rho=\rho^{\star}(b_0)$ (where 
 $\mu(\rho^{\star})=2b_0$). Note that at $\rho=\rho_h$,
 $\omega=\nu(\rho_h)=\omega_h$, and so this is the Hopf bifurcation to
 the Pyragus orbit.

The solutions for $\tau$ along the Hopf curve solve
 \[\cos(\omega\tau)=1-\frac{\mu(\rho)}{b_0}.\] 
Since $\mu(\rho_h)=0$, at $\rho=\rho_h$, $\omega\tau=2\pi n$, and
 similarly since $\mu(\rho^{\star})=2b_0$, at
 $\rho=\rho^{\star}$, $\omega\tau=\pi(2n+1)$.
The curve of Hopf bifurcations is thus tangent to the lines
 $\rho=\rho_h$ and $\rho=\rho^{\star}$, and forms a series of wiggles between
 these value of $\rho$. In particular, the curve is tangent to
 $\rho=\rho_h$ at $(\rho,\tau)=(\rho_h,\tau_h$), where
 $\omega=\omega_h$, the Hopf bifurcation to the Pyragus orbit.

\begin{figure}
\psfrag{rho}{\raisebox{-0.2cm}{$\rho$}}
\psfrag{b0}{$b_0$}
\begin{center}
\epsfig{file=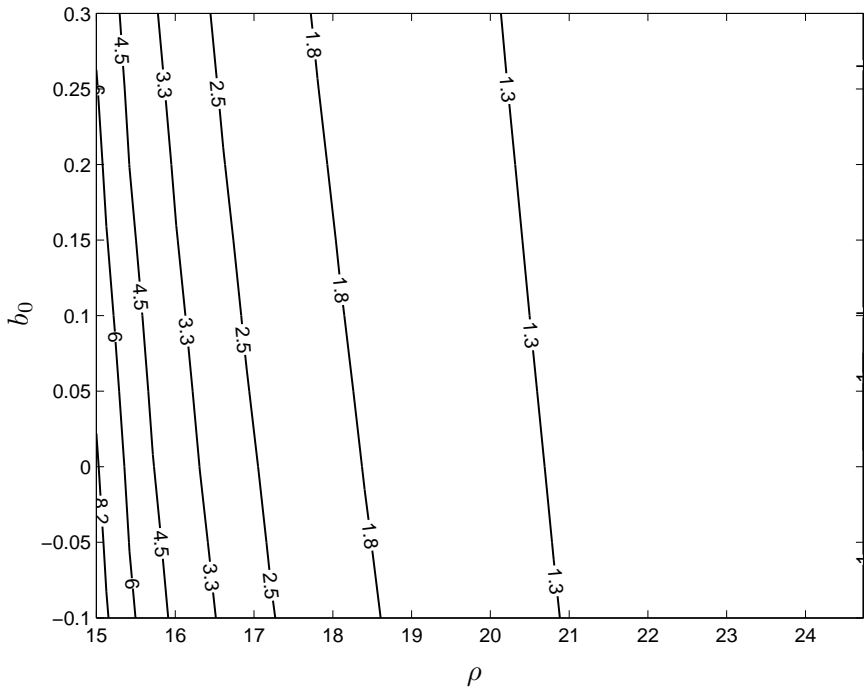, width=0.8\textwidth}
\end{center}
\caption{\label{fig:postab}The figure shows a contour plot of the
  magnitude of the largest Floquet
  multipliers for the Pyragus periodic orbit, as $\rho$ and $b_0$ are varied. The periodic orbit is
  unstable for all parameters shown. This figure was produced using DDE-BIFTOOL.}
\end{figure}

Therefore, for any value of $b_0$, the Pyragus curve
$\tau=\tau_P(\rho)$ (which originates at
$(\rho,\tau)=(\rho_h,\tau_h)$, and is also shown in
figure~\ref{fig:hopf_id}), will always be to the left of the Hopf
curves at 
$\rho=\rho_h$. The zero equilibrium is therefore stable
along the Pyragus curve close to the bifurcation point in
$\rho<\rho_h$, and so the Pyragus orbit will always bifurcate
unstably.
Figure~\ref{fig:postab}
shows a contour plot of the largest Floquet multipliers of the
periodic orbit, as $\rho$ and $b_0$ are varied. For all values shown the periodic orbit is unstable.


\section{Discussion}
\label{sec:conc}

In this paper, we have demonstrated how the mechanism used by
Fiedler~\etal~\cite{Fie06} for stabilizing periodic orbits with one unstable
Floquet multiplier carries over to higher dimensional systems, using
the Lorenz equations as an example. We use the results from their idealized example to inform our choice of the
feedback gain matrix, and this method follows
a set prescription which we expect could be used on other systems. First we
find the two-dimensional linear center eigenspace of the system with no
feedback at the Hopf bifurcation point. Feedback is then added in
the directions lying tangent to this center subspace, using a $2\times 2$ gain matrix of the form given by
Fiedler~\etal This then leaves only the two parameters $\beta$ and
$b_0$ to be chosen. For the example of the Lorenz equation, the
subcritical orbits are stabilized over a wide range of parameters. We
contrast this with an example of choosing the gain matrix as a real
multiple of the identity. In this case the Pyragus orbit could not be stabilised.

Choosing the gain matrix for our Lorenz equations example required a knowledge of the
linearization of the system at the bifurcation point.
This method may also be applicable in systems for which the governing
equations are not known, if it is possible to access perturbations to
the equilibrium solution near the Hopf bifurcation point, and hence extract the unstable
eigenvectors numerically from experimental data.

We have additionally shown that the Lorenz equations example contains
a codimension-two point which is also present in the normal form
example of~\cite{Fie06}. This double-Hopf point is not
generic. In the normal form example of~\cite{Fie06} there is an additional $SO(2)$ symmetry which is not
present in the Lorenz example.
Additional structures in the problem force a normally
codimension-three phenomena~\cite{GKL90} to be codimension-two, since
the frequencies of the bifurcating periodic orbits are forced to be in one:one
resonance at the codimension-two point. It would be of interest to
examine this degeneracy in more detail, by understanding the
mathematics behind the structure of the Hopf-Hopf bifurcation in these
examples. We intend to investigate further examples to see how
robust this bifurcation structure is, for example, whether it appears in say, the
Hodgkin--Huxley~\cite{GW00} or Belousov--Zhabotinsky~\cite{IHS97} examples.

\section*{Acknowledgements}
The authors would like to thank Luis Mier-y-Teran and David Barton for assistance with
DDE-BIFTOOL. We are grateful to an anonymous referee for several helpful and detailed suggestions.
This research was funded by NSF grant DMS-0309667.

\end{document}